\documentclass[journal]{IEEEtran}
\usepackage{cite}
\usepackage[numbers,sort&compress]{natbib}
\ifCLASSINFOpdf
\else
\fi
\usepackage{graphicx}		% to insert figures
\usepackage{amsmath}
\usepackage{amssymb}
\usepackage{amsfonts}
\usepackage{color}
\usepackage{xparse,xcolor}

\begin{document}
\title{Enhanced phase noise reduction in localized two-way optical frequency comparison}
%\title{Suppressing the fiber delay-limited phase noise in localized two-way optical frequency comparison}

\author{Long Wang, Ruimin Xue,  Wenhai Jiao,  Liang  Hu,~\IEEEmembership{Member,~IEEE,} Jianping Chen, and Guiling  Wu,~\IEEEmembership{Member,~IEEE}% <-this % stops a space
	\thanks{Manuscript received xxx xxx, xxx; revised xxx xxx, xxx. This work was 
		supported by the National Natural Science Foundation of China (NSFC) (62120106010, 61905143), the Zhejiang provincial Key Research and Development Program of China (2022C01156) and the National Science Foundation of Shanghai (22ZR1430200). (Corresponding author: Liang Hu; Guiling Wu)}
	\thanks{L. Wang, and R. Xue are with  the State Key Laboratory of Advanced Optical Communication Systems and  Networks, Department of Electronic Engineering, Shanghai Jiao Tong University,  Shanghai 200240, China (e-mail: long$_{-}$sjtu@sjtu.edu.cn; rmxue96@sjtu.edu.cns)} 
	\thanks{L. Hu, J. Chen, and G. Wu are with  the State Key Laboratory of Advanced Optical Communication Systems and  Networks, Department of Electronic Engineering, Shanghai Jiao Tong University,  Shanghai 200240, China, and also with the SJTU-Pinghu Institute of Intelligent Optoelectronics, Jiaxing Time-Transfer Photoelectric Co., Ltd., Pinghu 314200, China, China (e-mail: liang.hu@sjtu.edu.cn;  jpchen62@sjtu.edu.cn; wuguiling@sjtu.edu.cn)} 
	\thanks{Wenhai Jiao is with the Beijing Institute of Tracking and Telecommunication  Technology, Beijing 100094, China (email: jiaowh0927@163.com)}
	
	%\thanks{J. Shen is with the College of Physics and electronic information Engineering, Zhejiang Normal University, Jinhua, 321004, China (e-mail: shenjianguo@zjnu.cn).}
}
 

\markboth{Journal of Lightwave Technology,~Vol.~xxx, No.~xxx, XXX~2022}%
{Shell \MakeLowercase{\textit{et al.}}: Bare Demo of IEEEtran.cls for IEEE Journals}



\maketitle

\begin{abstract}
High-stability optical frequency comparison over fiber link enables the establishment of ultrastable optical clock networks, having the potential to promote a series of applications, including metrology, geodesy, and astronomy. In this article, we theoretically analyze and experimentally demonstrate a time-delayed local two-way (TD-LTW) optical frequency comparison scheme with improved comparison stability, showing that the fractional instability of optical frequency comparison over a 50-km transfer link can be reduced from $1.30\times10^{-15}$ to $5.25\times10^{-16}$ at the 1 s integration time with an improvement factor of 2.48. Additionally, we also for the first time model and experimentally verify the effect of the inhomogeneous phase noise along the fiber link on the system performance. We believe that the theory and technique proposed here will be helpful in developing the high-stability optical clock networks over large-area fiber links.
\end{abstract}

% Note that keywords are not normally used for peer review papers.

% \fi
\IEEEpeerreviewmaketitle

%%%%%%%%%%%%%%%%%%%%%%%%%%  body  %%%%%%%%%%%%%%%%%%%%%%%%%%
\section{Introduction}
\IEEEPARstart{W}{ith} the accuracy of optical clocks achieving the level of 
$10^{-18}$ \cite{ushijima2015cryogenic, brewer2019al+,nicholson2015systematic}, 
the large-scale, high-stability optical clock network is of increasing interest 
for many applications,  including radio astronomy 
\cite{Wang2015Square, 
		He2018Long}, clock-based geodesy \cite{2018Geodesy}, very long 
	baseline interferometry \cite{Clivati2017A, Cecilia2015A}, and 
	gravitational-wave detection \cite{Kolkowitz2016Gravitational, 
	Hu2017Atom}. To develop clock networks, fiber-based optical frequency 
	transfer and comparison 
are one of the most important ingredients \cite{Ma1994Delivering, 
foreman2007coherent, K2008Long, P2008High, Fabien2009High, G2009Optical, 
Predehl2012A, 2014Optical, 2020Passive1, 2020Passive2, Calosso2014Frequency1, 
Anthony2014Two, Stefani2015Tackling, 2017Hybrid}. Optical frequency transfer 
gives the possibility to distribute the frequency standard to different user 
sites \cite{Ma1994Delivering, foreman2007coherent, K2008Long, P2008High, 
Fabien2009High, G2009Optical, Predehl2012A, 2014Optical, 2020Passive1, 
2020Passive2} whereas optical frequency comparison compares the frequency 
standards \cite{Calosso2014Frequency1, Anthony2014Two, Stefani2015Tackling, 
2017Hybrid} between a wide number of metrology laboratories. 

The basic setups of optical frequency transfer and comparison are both constructed by an imbalanced Michelson interferometer, whose long arm is connected with the fiber link and short arm is spliced with a Faraday rotating mirror (FRM) to provide a local frequency reference \cite{Ma1994Delivering}. For optical frequency transfer, the local optical signal is delivered to the remote end along the transfer link, and then part of the beam is reflected back to the local site. By comparing the round-trip signal with the local frequency reference, the link-induced phase noise can be detected, and then the active noise compensation (ANC) based phase locked loop (PLL) is employed to compensate the phase noise induced by the fiber link. However, the finite propagation velocity of the optical signal limits the compensation performance, leaving a named delay-unsuppressed residual phase noise. The power spectral density (PSD) of this phase noise was first studied in \cite{P2008High}, whose expression for the ANC scheme at the offset frequency $f\leq 1/(4\tau)$ is,
\begin{equation}
 S_{\rm r}(\omega) \simeq\frac{1}{3}(2\pi f\tau)^2S_0(\omega)
 \label{eq1}
 \end{equation}
where $\tau$ is the one-way propagation delay of the optical fiber link \cite{P2008High}, and $S_0(\omega)$ is the phase noise PSD accumulated by the single-trip optical signal.

Taking into account the quadratic 
relationship between the phase noise PSD and the propagation delay as 
illustrated in Eq. (\ref{eq1}), Lopez and co-workers proposed a 
cascaded optical frequency transfer \cite{Lopez2010Cascaded} technique by 
dividing the whole transfer link into $N$ sub-links, theoretically resulting in 
a reduction of the phase noise PSD with a factor of $N$. However, there are 
several other factors practically limit the stability, including the 
out-of-band phase noise induced by the laser locking in the cascaded repeaters. Consequently, the long-term stability could be constrained by the noise floor \cite{chiodo2015cascaded}. In 2015, Calosso \emph {et al}  proposed a post-processing approach for optical frequency transfer, which can  overcome the classical limit of the delay-unsuppressed residual phase noise  \cite{C2015Doppler}. This scheme is realized by a delay subtraction algorithm  between the optical signal delivered at the remote end and the round-trip  signal detected at the local site. However, it is difficult to put this  proof-of-principle study into practice because the separated data acquisitions  at the different ends of the fiber link limits the real-time feedback control.  On one hand, if the remote signal is sent back to the local site and a feedback  servo is implemented, the residual phase noise will be determined by Eq. (\ref{eq1}) again. On the other hand, if the local beating note is sent to the remote site to accomplish feedback, it is hardly to generate a time delay for  the transferred optical frequency signal without introducing additional phase noise. In our pioneering work, a method of phase noise compensation based on  feed-forward control is proposed for the optical frequency  transfer \cite{2020Passive2, 2020Passive1, hu2021all, hu2020multi}. This scheme can  effectively reduce the system complexity and remove the servo bump induced by  the PLL module, but the residual phase noise PSD presents a factor  of 7 higher than that of Eq. (\ref{eq1}), meaning a worse frequency transfer  stability compared with the ANC scheme.

In the field of optical frequency comparison, the lower bound of 
the phase noise 
suppression defined by Eq. 
(\ref{eq1}) can be reduced in the two-way optical frequency comparison (TWC) 
scheme \cite{Calosso2014Frequency1, Anthony2014Two, Stefani2015Tackling, 
2017Hybrid}, in which the optical frequency signals located at two distant 
sites are delivered to the opposite ends, and each site detects the beating 
note between the local reference and received signals. Under the assumption 
that the link-induced phase noises are reciprocal for two propagation 
directions, the phase noise can be efficiently suppressed by the post 
processing \cite{Calosso2014Frequency1}. Theoretically, the TWC method provides 
a factor of 4 reduction compared to  Eq. (\ref{eq1}), meaning a lower 
delay-unsuppressed residual phase noise \cite{P2008High}. However, in the 
post-processing, the timebase mismatch of the data acquisition at two distant 
sites will sacrifice the acquired stability improvement in the TWC technique. 
For example, when the mismatch is larger than the one-way propagation time  
such as about 500 $\mu$s for 100-km fiber link, it will spoil the improvement 
achieved by the TWC method \cite{2017Effect}. To deal with the time mismatch of 
the data acquisition process and enable a real-time measurement, Lee \textit 
{et al} proposed a local two-way (LTW) frequency comparison method 
\cite{2017Hybrid}. Different from the TWC method, the ultrastable optical 
frequency signal arriving at the remote site is reflected back to the local 
site, passing through fiber link twice. Under this circumstance, the relative 
frequency of the two optical clocks can be obtained at the local site by 
comparing the frequency of the round-trip signal and the single-trip remote 
optical signal, eliminating the risk of the timebase mismatch caused by the 
separated data acquisitions. Nevertheless, its capability of noise suppression 
is still determined by the classical limit expressed in the Eq. (\ref{eq1}), 
illustrating a worse comparison stability than the well-worked TWC scheme.

In this paper, we propose a time-delayed local two-way comparison (TD-LTW) technique, enabling a real-time optical frequency comparison by simple frequency mixing at the local site and more importantly breaking the classical phase-noise limit expressed by the Eq. (\ref{eq1}) in the conventional LTW scheme \cite{2017Hybrid}. Specifically, the following sections are organized as below: Section II presents the theoretical foundation of the proposed TD-LTW scheme, and to the best of our knowledge, the effect of the inhomogeneous phase perturbation on the fiber link is first studied. Section III demonstrates the experimental setup and presents the main experimental results, including the comparison between the proposed TD-LTW technique and the conventional LTW scheme, and the investigation of the inhomogeneous phase noise perturbation on the fiber link. Finally, Section \ref{sec4} gives the discussion and conclusion. 

\section{Theoretical analysis on the delay-unsuppressed residual phase noise}
\label{sec2}
\subsection{Theoretical foundation of the proposed TD-LTW scheme}

\begin{figure*}[h!]%[htbp]
\centering
%\fbox{\includegraphics[width=\linewidth]{fig5}}
\includegraphics[width=1\linewidth]{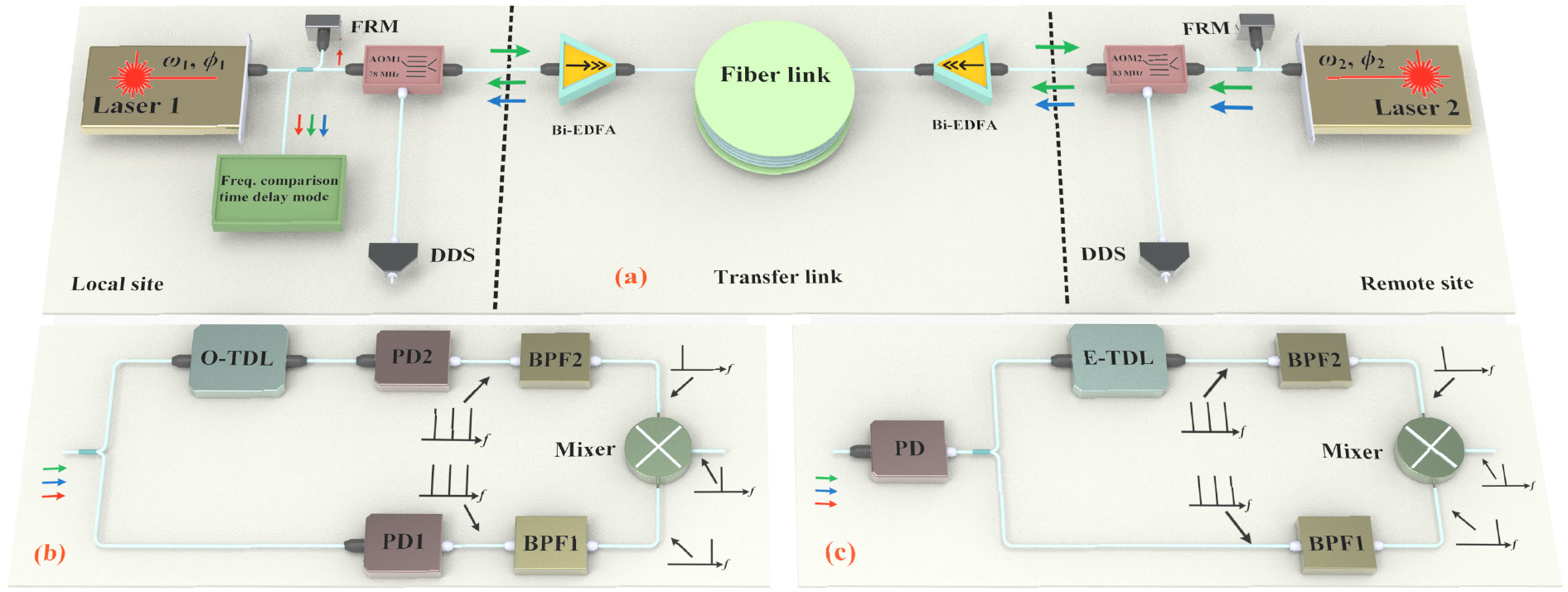}
\caption{The schematic diagram of the proposed TD-LTW optical frequency comparison (a), in which the time delay can be implemented in the optical domain (b) or the electrical domain (c).  By removing the time delay line, i.e. setting $\delta \tau=0$, the configuration can function as the conventional LTW scheme. FRM: Faraday rotating mirror, AOM: acousto-optic modulator, Bi-EDFA: bidirectional Erbium-doped optical fiber amplifier, DDS: direct digital synthesizer, PD: photo-detector, O-TDL: optical tunable delay line, E-TDL: electrical tunable delay line, BPF: band-pass filter.}
\label{fig1}
\end{figure*}

Figure \ref{fig1} shows the schematic diagram of the proposed TD-LTW optical frequency comparison technique. Two distant sites, both equipped with ultrastable 
lasers, are connected by an optical fiber link. At the local site, the output of the ultrastable laser 1 is split into two parts by an optical coupler, where one indicted by the red arrows is reflected by the local FRM and then falls into the time-delay frequency comparison module, serving as the reference beam. The other part, shown by the green arrows, is transferred towards the remote site via the optical fiber link. At the remote  site, there are two optical frequency signals being injected into the same fiber link  and transferred toward the local site. One comes from the output of the ultrastable laser 2, indicted by the blue arrow, and  the other beam (green arrow) coming from the local site is  reflected by the remote FRM. After the propagating along the fiber link, these optical frequency signals are both collected by 
the time-delayed frequency comparison module. Furthermore, in this setup, two acousto-optic modulators (AOMs) are placed at the ends of the fiber link to remove the influence of the backscattering noise and other unwanted reflections \cite{Ma1994Delivering}. By this means, three desired optical signals, the reference beam, the round-trip beam and the single-trip signal coming from the laser 2, are simultaneously received by the time-delayed frequency comparison module.

To realize the time-delayed frequency comparison, the time delay can be implemented in the optical domain (e.g., fiber) or electrical domain, as shown in Fig. \ref{fig1} (b) and (c) respectively. In terms of the optical time delay method, the received signals are split into two unbalanced detection paths, which 
have a relative delay $\delta \tau$ generated by the optical delay line. Then, the relative frequency between the two distant ultrastable lasers can be obtained 
by mixing the two filtered beating signals detected by photo-detectors (PDs). Alternatively, the electrical time delay method implemented in the FPGA platform can be adopted \cite{won2016time} as schematically illustrated in Fig. 1 (c). In this method, the optical signals are first converted into the microwave signals, and then are divided into two paths. One branch is directly filtered by a band-pass filter and injected into the mixer. The other branch, after passing through the electrical time delay module and a band-pass filter, is also injected into the mixer to accomplish the frequency comparison. 

Here we choose the first comparison method to deduce the TD-LTW method proposed, and the same results can also be obtained for the electrical delay method by following the same procedure.  By properly choosing the driving frequencies of the AOMs ($\Omega_1$ for AOM1 and $\Omega_2$ for AOM2),  three beating signals recovered by the PD1 can be expressed as,
\begin{equation}
E_{\rm PD1,i}\propto\cos \left[\Delta \omega_{i}t+\phi_{i}(t)  \right], i=1,2,3,
\label{eq2}
\end{equation}
where $\Delta \omega_{i}$ are the frequencies of the beating signals, being equal to $\Delta \omega_{1}=2(\Omega_1+\Omega_2)$, $\Delta \omega_{2}=\omega_1-\omega_2-\Omega_1-\Omega_2$, and $\Delta \omega_{3}=\Omega_1+\Omega_2+\omega_1-\omega_2$, and  $\phi_{1}=\phi_{\rm RT}(t)$, $\phi_{2}=\Delta \phi-\phi_{\rm R-L}(t)$, $\phi_{3}=\phi_{\rm L-R}(t)+\Delta \phi$ with  $\Delta \phi=\phi_1-\phi_2$ being the relative phase between the laser 1 and the laser 2, $\phi_{\rm RT}(t)$ being the fiber-induced phase noise experienced by the round-trip signal, and $\phi_{\rm R-L}(t)$ ($\phi_{\rm L-R}(t)$) being the fiber-induced phase noise experienced by the optical frequency signal propagating from the remote (local) site to the local (remote) site. Moreover, $\omega_1$ and $\omega_2$ are the output frequencies of laser 1 and laser 2, respectively. 

Likewise, the beating signals recovered  by the PD2 can be written as,
\begin{equation}
E_{\rm PD2,i}\propto\cos \left[\Delta \omega_{i}t+\phi_{i}(t-\delta \tau)  \right],
\label{eq2}
\end{equation}
where $\delta\tau$ is the time delay induced by the optical delay fiber. 

By properly setting the frequencies of the band-pass filters (BPF1 and BPF2), different beating signals can be selected to conduct frequency mixing to suppressing the link-induced phase noise, and then the mixing output can be used to extract the relative phase $\Delta \phi$. Here, three kinds of mixing signals can be used to evaluate the relative phase $\Delta \phi$, which can be expressed as,
\begin{equation}
	E_1(t)\propto \cos \left[(\Delta \omega_{2}+\Delta \omega_{1}/2)t+\Delta\phi_{\rm L1}(t)+\Delta \phi  \right],
\end{equation}
\begin{equation}
	E_2(t)\propto \cos \left[(\Delta \omega_{3}-\Delta \omega_{1}/2)t+\Delta\phi_{\rm L2}(t)+\Delta \phi  \right],
\end{equation}
\begin{equation}
	E_3(t)\propto \cos \left[(\Delta \omega_{3}-\Delta \omega_{2})t/2+\Delta\phi_{\rm L3}(t)+\Delta \phi  \right],
\end{equation}
respectively. The phase terms, $\Delta\phi_{\rm L1}(t)=\phi_{\rm RT}(t)/2-\phi_{\rm R-L}(t-\delta\tau)$, $\Delta\phi_{\rm L2}(t)=\phi_{\rm L-R}(t)-\phi_{\rm RT}(t-\delta\tau)/2$, and $\Delta\phi_{\rm L3}(t)=[\phi_{\rm L-R}(t)-\phi_{\rm R-L}(t-\delta\tau)]/2$ are the residual phase noises induced by the optical fiber link, which affect the evaluation of the $\Delta \phi$.

Following the same procedure presented in \cite{P2008High}, the accumulated phase noise of the round-trip and single-trip signal can be written as,
\begin{equation}
	\begin{aligned}
	\phi_{\rm RT}(t)&=\int_{0}^{L}h(z)\delta\phi\left[z,t-(2\tau-z/c_n)\right]{\rm d}z\\
	                &+\int_{0}^{L}h(z)\delta\phi\left(z,t-z/c_n\right){\rm d}z
	\end{aligned}  
\end{equation}
\begin{equation}
	\phi_{\rm R-L}(t)=\int_{0}^{L}h(z)\delta\phi\left(z,t-z/c_n\right){\rm d}z
\end{equation}
\begin{equation}
	\phi_{\rm L-R}(t)=\int_{0}^{L}h(z)\delta\phi\left[z,t-(2\tau-z/c_n)\right]{\rm d}z
\end{equation}   
where $\delta\phi(z,t)$ is the phase perturbation on the fiber at time $t$ and position $z$, $\tau=L/c_n$ is the propagation delay along the fiber, and $c_n$ is the velocity of the optical frequency signal in the fiber link. Note that a spatial distribution function $h(z)$ is used to describe the inhomogeneous phase perturbation on the fiber link. By this means, the residual phase noises in Eq. (4)-(6) can be expressed as,
\begin{equation}
	\Delta\phi_{\rm L1}(t)=\int_{0}^{L}h(z)(z/c_n-\tau+\delta\tau)\delta\phi'(z,t){\rm d}z,
\end{equation}
\begin{equation}
	\Delta\phi_{\rm L2}(t)=\int_{0}^{L}h(z)(z/c_n-\tau+\delta\tau)\delta\phi'(z,t){\rm d}z,
\end{equation}
\begin{equation}
	\Delta\phi_{\rm L3}(t)=\int_{0}^{L}h(z)(z/c_n-\tau+\delta\tau/2)\delta\phi'(z,t){\rm d}z,
\end{equation}
where the first-order Taylor approximation of $\delta\phi(z,t)$ is used and $\delta\phi'(z,t)$ represents the first-order derivative of the phase perturbation $\delta\phi(z,t)$. 

Under the assumption that the link-induced phase noise is uncorrelated with position, the residual phase noise PSDs of the TD-LTW optical frequency comparison scheme using the different evaluation signals in Eq. (4)-(6) are,
\begin{equation}
S_{\rm L1}(\omega,\delta\tau)={\omega^2S_0(\omega)}\int_{0}^{L}\frac{h^2(z)}{L}(z/c_n-\tau+\delta\tau)^2{\rm d}z,
\label{eq8_1}
\end{equation}
\begin{equation}
S_{\rm L2}(\omega,\delta\tau)={\omega^2S_0(\omega)}\int_{0}^{L}\frac{h^2(z)}{L}(z/c_n-\tau+\delta\tau)^2{\rm d}z,
\label{eq9_1}
\end{equation}
\begin{equation}
S_{\rm L3}(\omega,\delta\tau)={\omega^2S_0(\omega)}\int_{0}^{L}\frac{h^2(z)}{L}(z/c_n-\tau+\frac{\delta\tau}{2})^2{\rm d}z,
\label{eq10_1}
\end{equation}
respectively. If the perturbation on the fiber link is homogeneous (i.e. $h(z)=1$) and uncorrelated in position, we can get,
\begin{equation}
    S_{\rm L1}(\omega,\delta\tau)=\frac{\omega^2}{3\tau}\left[(\delta\tau)^3-(\delta\tau-\tau)^3\right]S_0(\omega),
    \label{eq8}
\end{equation}
\begin{equation}
    S_{\rm L2}(\omega,\delta\tau)=\frac{\omega^2}{3\tau}\left[(\delta\tau)^3-(\delta\tau-\tau)^3\right]S_0(\omega),
    \label{eq9}
\end{equation}
\begin{equation}
    S_{\rm L3}(\omega,\delta\tau)=\frac{\omega^2}{3\tau}\left[(\frac{\delta\tau}{2})^3-(\frac{\delta\tau}{2}-\tau)^3\right]S_0(\omega).
    \label{eq10}
\end{equation}

Once $\delta\tau=0$,  the residual phase noise PSDs of Eq. (\ref{eq8})-(\ref{eq10}) become,
\begin{equation}
 S_{\rm LTW}(\omega)=\frac{1}{3}(2\pi f\tau)^2S_0(\omega),
 \label{eq11}
 \end{equation}
which is the typical residual phase noise PSD acquired by the conventional LTW optical frequency comparison scheme \cite{2017Hybrid}. Alternatively, by choosing the optimal time delay $\delta \tau$, the residual phase noise PSDs of Eqs. (\ref{eq8})-(\ref{eq10}) become,
 \begin{equation}
    S_{\rm TD-LTW}(\omega, \delta\tau=\tau )=\frac{1}{12}(2\pi f\tau)^2S_0(\omega),
    \label{eq12}
\end{equation}
where the optimal values for Eqs. (\ref{eq8})-(\ref{eq10}) are $\delta\tau=\tau/2$, $\delta\tau=\tau/2$, and $\delta\tau=\tau$, respectively. 

Here, we define an improvement factor as the Allan deviation (ADEV) ratio of the LTW scheme to that of the proposed TD-LTW scheme. According to Eq. (19) and (20), it can be inferred that the further suppression of the residual phase noise by a factor of 4 can result in an improvement factor of 2 in the ADEV-expressed fractional frequency stability under the assumption of the homogeneous perturbation. Moreover, compared with Eqs. (16) and (17), we can be seen that there is a denominator 2 in the term of $\delta \tau$ in Eq. (18), indicating a smaller influence resulting from the deviation from the optimal time delay value. Therefore, the method given by Eq. (6) will be adopted in the following experiments.

\subsection{Inhomogeneous phase perturbation induced by the fiber link}

\begin{figure}[h!]
	\centering
	\includegraphics[width=1\linewidth]{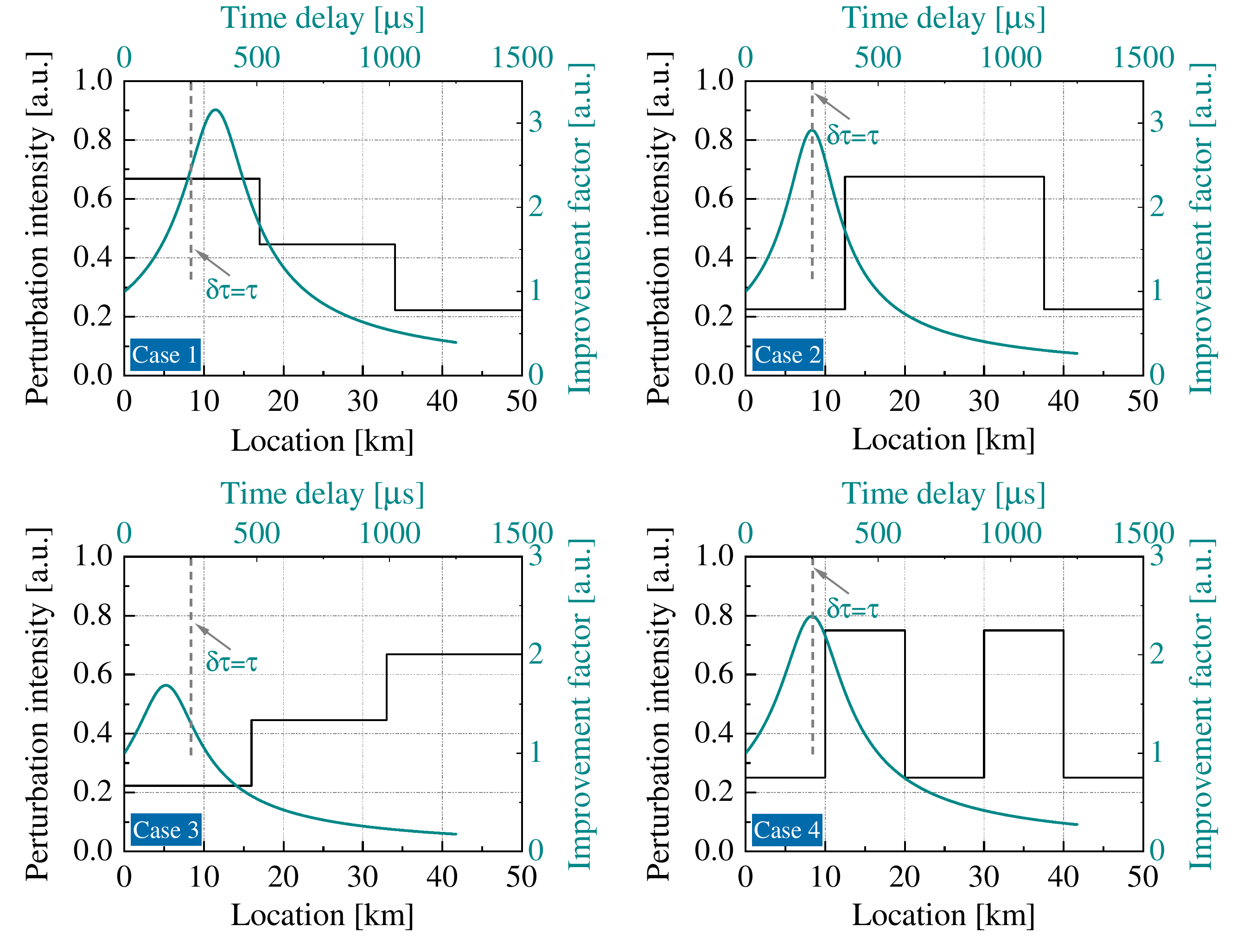}
	\caption{The distribution functions of four kinds of piecewise uniform functions with the same total integrated phase noise (left and bottom axes) and the improvement factors achieved by the TD-LTW scheme under different time delay $\delta \tau$ (top and right axes). }
	\label{fig20}
\end{figure}

Considering the inhomogeneous phase noise perturbation along the fiber link, the behaviours of the phase noise suppression and the improvement achieved by the proposed TD-LTW technique will be different from the homogeneous case. Here, we assume that the phase noise perturbation distribution follows a piecewise uniform function, 
\begin{equation}
	h(z)=\sum_{i=1}^{N}h_i{\rm rect}\left(\frac{z-z_i}{z_{{\rm d},i}}\right),
\end{equation}
where ${\rm rect}[(z-z_i)/z_{{\rm d},i}]$ is equal 1 for $0\leq[(z-z_i)/z_{{\rm d},i}]\leq1$ and 0 otherwise. In the following simulation, four cases of the perturbation distribution are discussed, as shown in Fig. \ref{fig20}. 

Based on the phase noise perturbations with the same total integrated fiber phase noise shown in Fig. 2, the phase noise suppression ratios of different optical frequency comparison schemes are shown in Table 1. Here, the suppression ratios of the  LTW and TD-LTW schemes are obtained by calculating the integral terms ($\int_{0}^{L}\frac{h^2(z)}{L}(z/c_n-\tau+\frac{\delta\tau}{2})^2{\rm d}z$) in Eq. (15) with different time delay ($\delta\tau$). The ratio of the TWC scheme is obtained based on the Eq. (4) in \cite{Anthony2014Two} by taking the distribution function $h(z)$ into account. In the four inhomogeneous cases, all the integral factors of the conventional LTW scheme are higher than that of the proposed TD-LTW and TWC schemes, meaning a higher residual phase noise. The improvement factors of the four distributions are also shown in Table 1, demonstrating that the proposed TD-LTW technique can enhance the phase noise reduction under the inhomogeneous perturbation. 

Note that the improvement achieved by the TD-LTW scheme with $\delta \tau =\tau$ is better when the perturbation is concentrated on the front part of the fiber link. In particular, the best improvement can be achieved when the phase noise perturbation mainly locates at the center. Moreover, for the symmetric distribution in the case 1 and 3, the difference in the improvement factors comes from the LTW scheme, where the performances of the TD-LTW scheme in the phase noise suppression are same. More importantly, the optimal time delay $\delta \tau$ is different for the different phase noise perturbations. For example, $\delta \tau =\tau$ is only the optimal choice when the perturbation follows the symmetric distribution as demonstrated in the case 2 and 4 in Fig. 2. In the case 1 and 3, the optimal time delays are slight deviation from $\delta \tau =\tau$ and the optimal improvement factors achieved by the TD-LTW scheme in the case 1 and 3 can reach 3.16 and 1.69, respectively.

\begin{table}[]
	\caption{The comparison of different schemes under the inhomogeneous phase perturbations.}
	\vspace{1pt}
	\centering
	\begin{tabular}{ c  c  c  c  c }
		\hline
		Schemes & Case 1 & Case 2 & Case 3  & Case 4\\
		\hline
		LTW  & $7.72 \cdot  10^{-9}$ & $4.48 \cdot  10^{-9}$ & $2.21\cdot 10^{-9}$ & $4.97 \cdot  10^{-9}$           \\
		TD-LTW  & $1.29\cdot 10^{-9}$ & $5.27\cdot 10^{-10}$ & $1.29\cdot 10^{-9}$ & $8.67\cdot 10^{-10}$            \\
		TWC  & $1.29\cdot 10^{-9}$ & $5.27\cdot 10^{-10}$ & $1.29\cdot 10^{-9}$ & $8.67\cdot 10^{-10}$           \\   
		\hline       
		IF($\delta \tau =\tau$) & 2.45 & 2.92 & 1.31 & 2.38\\
		IF$_{\rm optimal}$ & 3.16 & 2.92 & 1.69 & 2.38\\
		\hline 
	\end{tabular}
	\label{bs2}
\end{table}

\section{Experiment setup and results}
\label{sec3}

\subsection{Experiment setup for investigating the proposed TD-LTW scheme}
To prove the stability improvement achieved by the TD-LTW scheme, a verification experiment is conducted based on the setup illustrated in Fig. 3. To effectively evaluate the proposed technique, here we use one laser to simulate two independent lasers at two distant sites and the frequency of the remote one is downshifted by the acoustic optical modulator 3 (AOM3) with the value of 38 MHz. To generate the time delay required in Eq. (18), a 50-km delay fiber is placed before the PD2. The AOM1 ($-1$ order) and AOM2 ($+1$ order) work at the frequency of $78$ MHz and 83 MHz, respectively. In this configuration, two beating signals at frequency of 43 MHz and 33 MHz are selected by the BPF1 and BPF2, respectively, each of which represents the single-pass signal. After frequency mixing the two filtered beating signals, the mixing output 76 MHz is divided by 2 and then used for evaluating the TD-LTW optical frequency comparison results. It is worth noting that considering the coefficient 40 ps/(K$\cdot$km) of the single-mode fiber, the artificial time delay variations of the delay fiber caused by the temperature fluctuations is negligible. For example, it will introduce $\sim$ 20 ns delay variation if the temperature fluctuations of 10 K over a 50 km time delay fiber. Moreover, to boost the attenuation coming from the 50-km time delay fiber we add an erbium-doped fiber amplifier (EDFA) before the PD2. 

\begin{figure}[h!]
\centering
\includegraphics[width=1\linewidth]{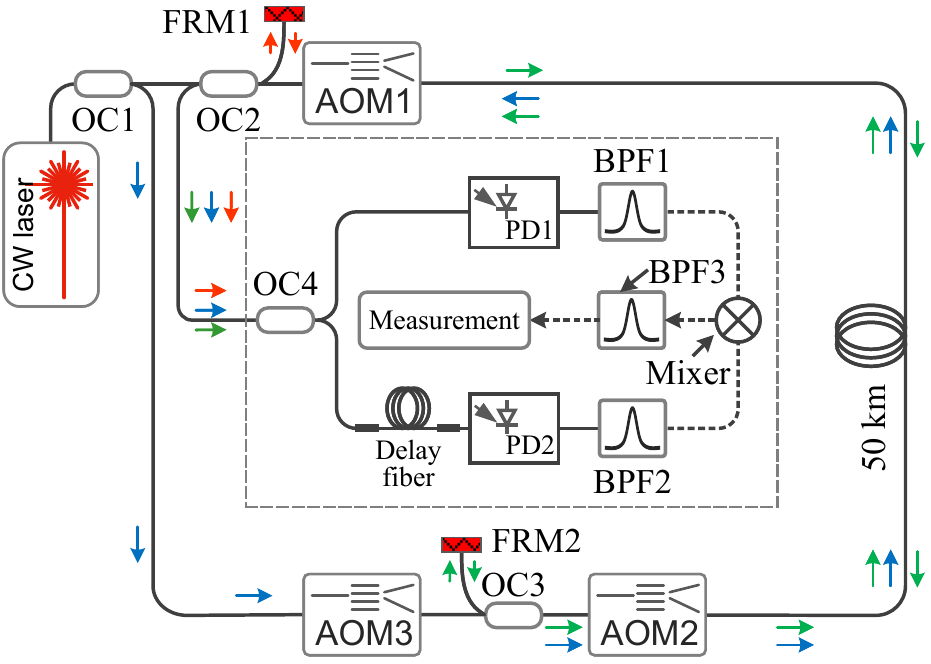}
\caption{The experimental setup for verifying the improvement achieved by the proposed TD-LTW optical frequency comparison scheme. One laser is employed to simulate two independent lasers at two sites, and the frequency of the remote one is downshifted by the AOM3 with the frequency of 38 MHz. OC: optical coupler, FRM: Faraday rotating mirror, AOM: acousto-optic modulator, PD: photo-detector, BPF: band-pass filter.}
\label{fig21}
\end{figure}

\subsection{The spectrum characterization of the optical frequency comparison}

\begin{figure}[h!]
	\centering
	\includegraphics[width=1\linewidth]{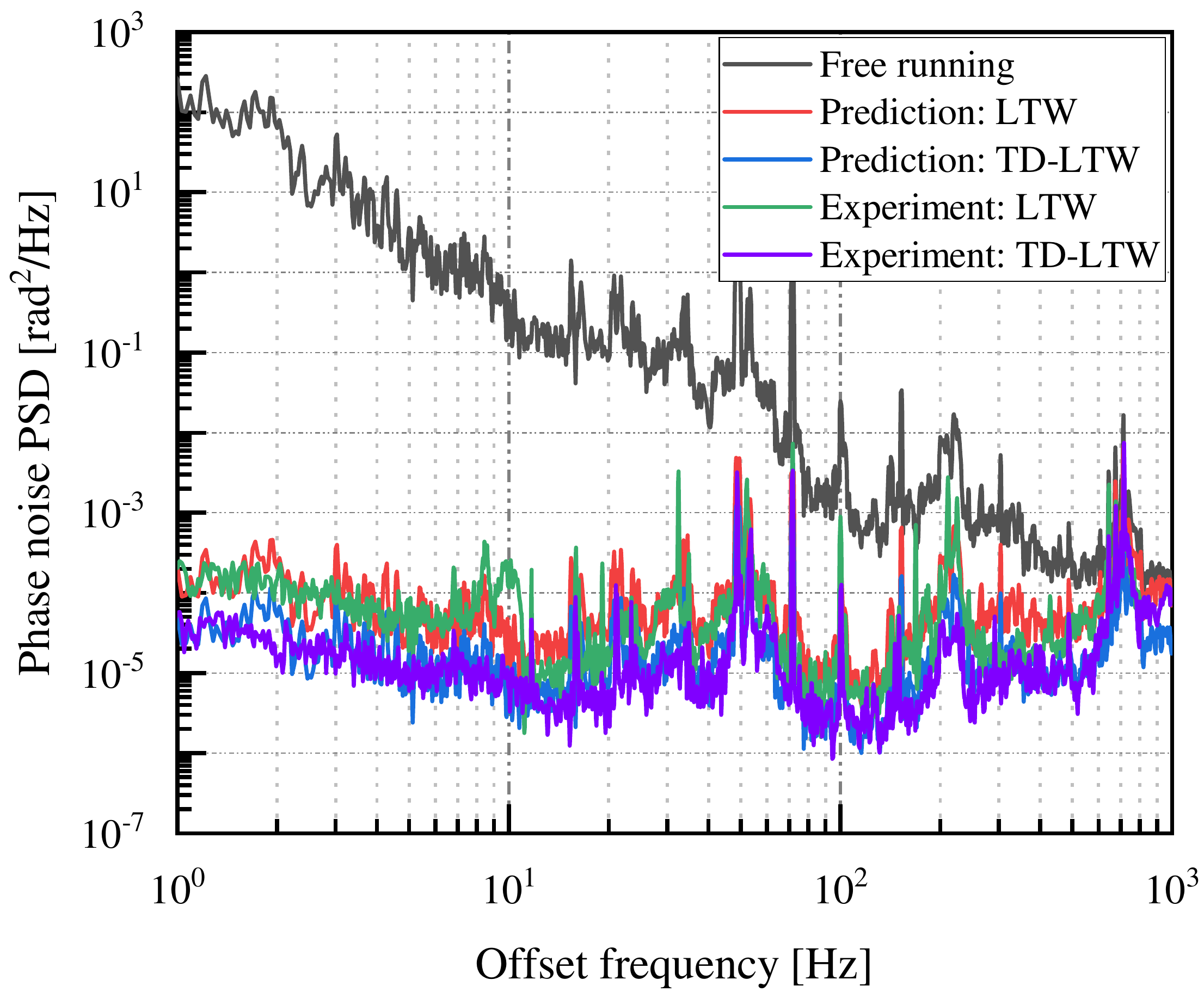}
	\caption{The measured phase noise PSDs of the 50-km free-running link (black 
	curve) and the phase noise suppressed link based on the LTW optical frequency comparison scheme (green curve) and the TD-LTW optical frequency comparison scheme (violet curve). The theoretical phase PSDs of the delay-unsuppressed residual phase noise predicted by Eq. \ref{eq11} (red 
	curve) and Eq. \ref{eq12} (blue curve) are also presented.}
	\label{fig3}
\end{figure}

The frequency-domain characterization of the optical frequency comparison expressed as the phase noise PSD is shown in Fig. \ref{fig3}. The phase noise PSD of the free-running fiber link (black line) evaluated by the 33 MHz beating signal approximately follows a power-law dependence of $S_{\phi}(f)=100f^{-2}$ for $f\leq 1$ kHz. By performing the proposed TD-LTW optical frequency comparison scheme, the residual phase noise PSD illustrates a $h_0f^0$ dependence, as drawn by the purple curve, which is almost consistent with the theoretical prediction as expressed by Eq. (\ref{eq12}). As a comparison, by removing the time delay fiber, the residual phase noise PSD of the conventional LTW scheme is also measured, as illustrated by the green curve. Its value is about four times higher than that of the TD-LTW scheme, demonstrating that an 6-dB enhancement the in phase noise reduction can be achieved by using the proposed TD-LTW optical frequency comparison scheme.

\begin{figure*}[h!]%[htbp]
\centering
%\fbox{\includegraphics[width=\linewidth]{fig7-1}}
\includegraphics[width=1\linewidth]{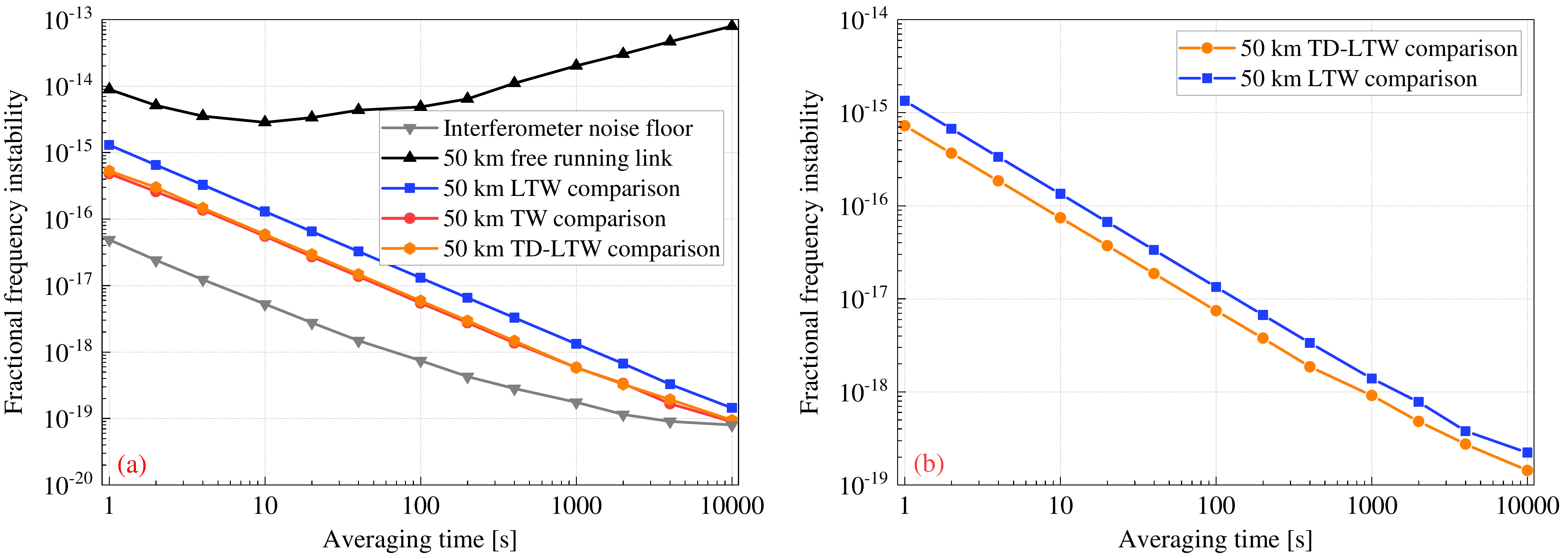}
\caption{The measured fractional frequency instabilities of optical frequency comparison using two different fiber spools. The black triangles, gray triangles, blue squares, orange circles, and red circles are the results of the 50-km free-running link, noise floor, LTW scheme, TD-LTW scheme, and TWC scheme, respectively. These measurements are conducted by a non-averaging ($\Pi$-type) frequency counter, expressed as the ADEV.}
\label{fig7-1}
\end{figure*}

\subsection{The stability and accuracy of optical frequency comparison}

To further investigate the instability of optical frequency comparison, we measured the fractional frequency instability characterized by the ADEV as shown in Fig. \ref{fig7-1} (a) and (b). In Fig. \ref{fig7-1} (a), it can be seen that the fractional frequency instability of the free running link (black triangle) fluctuates in the $10^{-14}$ level, which is not adequate for the current optical clocks. By removing the time delay fiber, the setup in Fig. 3 functions as the conventional LTW optical frequency comparison scheme. In this case, the ADEV decreases from $1.30\times10^{-15}$ at 1 s integration time to $1.44\times10^{-19}$ at 10,000 s with a slope of $\tau^{-1}$. After adding a 50 km time delay fiber, the ADEV scales down from $5.25\times10^{-16}$ at 1 s to $9.50\times10^{-20}$ at 10,000 s, which approaches the performance of the TWC scheme. The fractional frequency instability of the LTW and TD-LTW schemes using another fiber spool is shown in Fig. \ref{fig7-1} (b), whose values at 1 s integration time are $1.34\times10^{-15}$ and $7.25\times10^{-16}$, respectively. According to the measured ADEV, the experimental improvement factors for the two cases are 2.48 and 1.85, respectively.  It is important to stress that the improvement could be achieved by acquiring the frequency counter with the $\Lambda$-mode \cite{dawkins2007considerations}, which will be comprehensively investigated in our future work. The deviation from the theoretical improvement factor by Eq. (19) and (20) can be attributed to the inhomogeneous phase noise perturbation on the fiber link as discussed in the section \ref{sec2}. As a comparison, we also measured the noise floor by replacing the fiber link with a short fiber. We can clearly see that the long term stability is limited by $10^{-19}$, which is mainly due to the phase noise introduced by the out-of-loop path \cite{xue2021branching, hu2021performance}.

\begin{figure}[h!]
	\centering
	\includegraphics[width=1\linewidth]{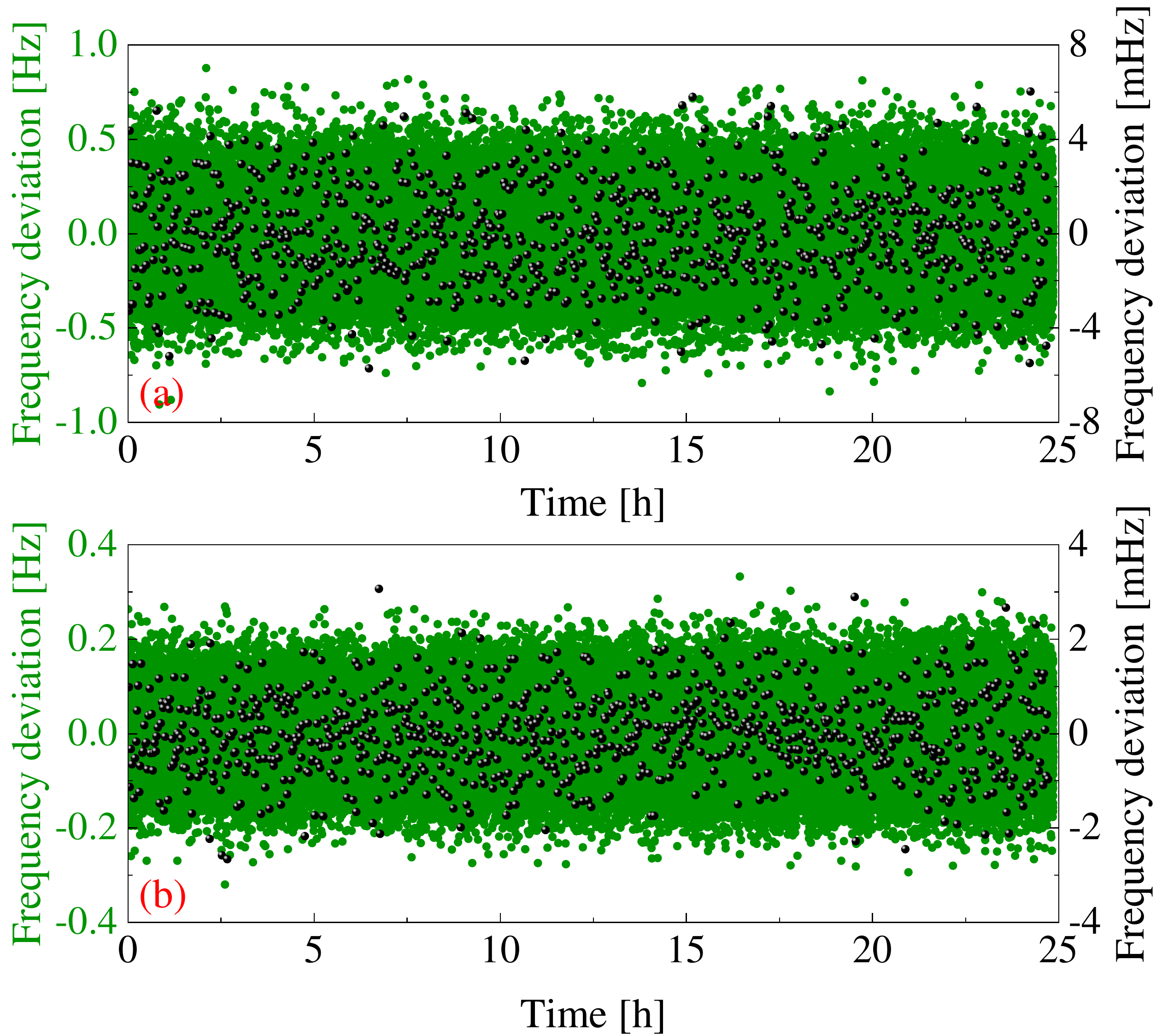}
	\caption{Frequency deviation of the optical frequency comparison. These results were measured by a dead-time free $\Pi$-type frequency counter at 1 s gate time over 25 hours (green points, left axis). The unweighted mean values of all all cycle-slip free 100 s intervals are shown by the black points (right axis).}
	\label{fig5-1}
\end{figure}

Considering that the accuracy of the frequency comparison can not be presented in the instability analysis, we measured the frequency deviation using an $\Pi$-type frequency counter, which worked at 1 s gate time. Here, we measured two spool fibers as used in the fractional frequency instability evaluation. For the first fiber spool, the frequency deviation of the LTW scheme over successive 90000 s was recorded, shown in Fig. \ref{fig5-1} (a) by the green points. By averaging the data every 100 seconds, 900 data points (black points, right frequency axis) are produced, as also shown in Fig. \ref{fig5-1} (a). The 900 data points have an arithmetic mean of 41.5 $\mu$Hz ($2.15 \times 10^{-19}$) and a standard deviation of 2.08 mHz ($1.08 \times 10^{-17}$). The frequency deviation of the TD-LTW scheme is shown in Fig. 6 (b), whose 900 averaged points have an arithmetic mean of 2.95 $\mu$Hz ($1.03 \times 10^{-20}$) and a standard deviation of 0.875 mHz ($4.04 \times 10^{-18}$). Thus, the statistical fractional frequency uncertainties of the conventional LTW and the TD-LTW optical frequency comparison schemes are $3.60 \times 10^{-19}$ and $1.35 \times 10^{-19}$, respectively. These values satisfy the requirement of the most accurate optical clocks to date. Following the same procedure, we also evaluate the frequency comparison accuracy of the LTW and TD-LTW schemes for the second fiber spool. The mean frequency offset and the standard deviation of 792 data points in the total 79200 seconds of the LTW scheme are 36.5 $\mu$Hz ($1.89\times 10^{-19}$) and 2.04 mHz ($1.05\times 10^{-17}$), respectively, and that of the TD-LTW scheme are 0.65 $\mu$Hz ($3.38\times 10^{-21}$) and 1.27 mHz ($0.65\times 10^{-17}$), respectively. Thus, the statistical fractional frequency uncertainties of the conventional LTW and the TD-LTW optical frequency comparison schemes are $3.73 \times 10^{-19}$ and $2.31 \times 10^{-19}$, respectively.

\begin{figure}[h!]
\centering
\includegraphics[width=1\linewidth]{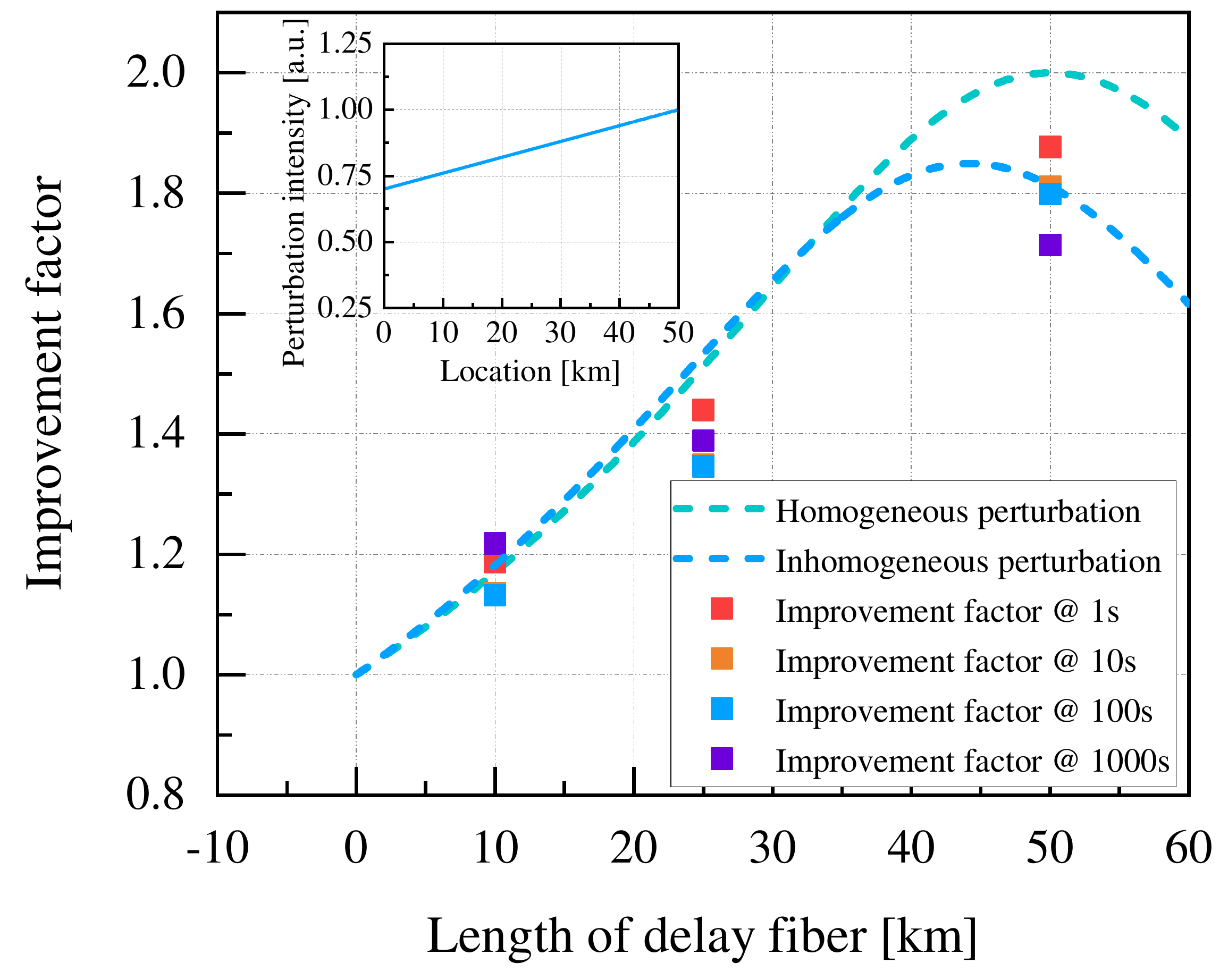}
\caption{The measured improvement factors achieved by the TD-LTW scheme. The data points here are all obtained by the second fiber spool. The green curve is calculated under the assumption of homogeneous  phase noise perturbation. The blue curve represents the theoretical improvement factor under the inhomogeneous phase noise perturbation, whose distribution follows the blue line in the inset.}
\label{fig5}
\end{figure}

To further analyze the dependence of the fractional frequency instability on the time delays $\delta\tau$, we measured the fractional frequency instabilities of the optical frequency comparison based on the TD-LTW optical frequency comparison scheme with the different time delay. The time delay was changed by changing the length of the delay fiber before the PD2. The improvement factors of the ADEV using the second fiber spool at the 1 s, 10 s, 100 s, and 1000 s integration time are drawn in Fig. 7. Moreover, the theoretical improvement factors under the homogeneous and inhomogeneous phase noise perturbations are plotted by the green and blue curves, respectively. In this case, it is clear that the 50-km delay fiber approaches the optimal value.

\subsection{The influence of the inhomogeneous phase noise perturbation}
To further investigate the influence of the inhomogeneous phase noise perturbation on the fractional frequency stability of the LTW and TD-LTW scheme, an artificial perturbation is applied to stimulate the inhomogeneous phase noise case. 
Here, two 25 km optical fiber spools were connected to form an 50 km fiber link, one of which was blown by an electric fan and the other was placed in a quiet position. By exchanging two ends of the fiber link, the phase noise distributions like the case 1 and 3 in Fig. 2 are experimentally generated, and the measured ADEV of the LTW and TD-LTW schemes is shown in Fig. 8. It can be seen that the instability of the LTW scheme is higher when the artificial phase noise perturbation is concentrated on the first span of the fiber link, which is consistent with the above theoretical analysis. Quantitatively, when the front 25 km fiber spool is blown by the electric fan, the ADEV of the LTW and TD-LTW schemes follows the dependence of $3.79\times 10^{-15}/\tau'$ and $1.58\times 10^{-15}/\tau'$, respectively. Here, $\tau' $ represents the averaging time. After exchanging the connection ends between the fiber link and the sites, the ADEV of the LTW and TD-LTW schemes follows the dependence of $2.78\times 10^{-15}/\tau'$ and $1.55\times 10^{-15}/\tau'$, respectively, demonstrating that the improvement factors of the two cases are 2.39 and 1.76, respectively. 

\begin{figure}[h!]
\centering
\includegraphics[width=1\linewidth]{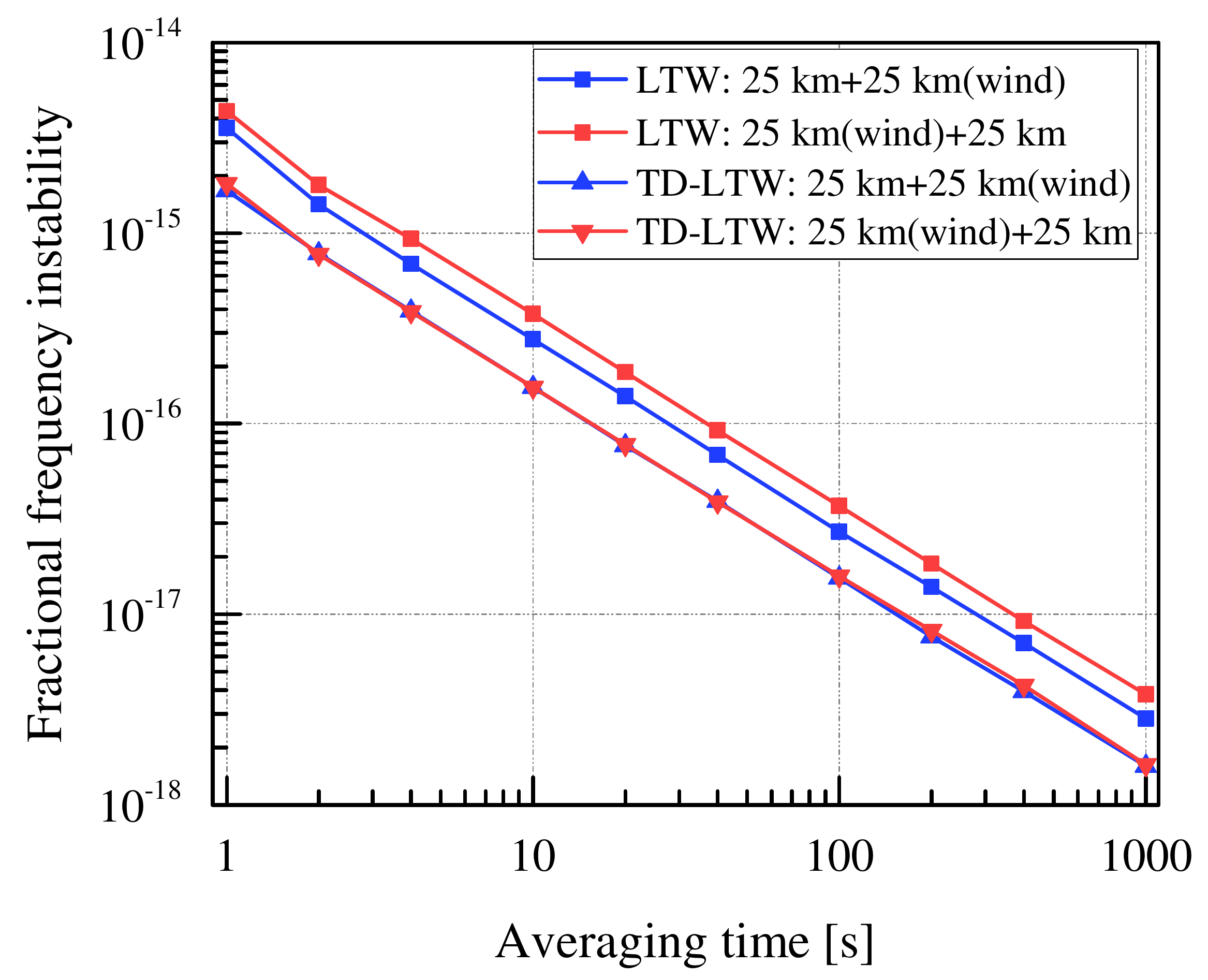}
\caption{The measured fractional instability of the optical frequency comparison under the artificial phase noise perturbation on the fiber link. The results presented by the blue squares and triangles are measured when the front 25 km fiber spool was blown by an electric fan and the other was placed in a quiet position. The red ones are obtained by exchanging the connection ends between the fiber link and the sites.}
\label{fig5}
\end{figure}

\section{Discussion and conclusion}
\label{sec4}

In section II, two configurations of the frequency comparison module are proposed to conduct the TD-LTW technique, based on the optical time delay or the electrical time delay, and the first one was adopted to generate the desired time delay in this work. For the second method, the configuration
adopted in \cite{Calosso2014Frequency1} is recommended, in which the phase variation is converted into the voltage variation, sampled and processed by a digital circuit. We think the two methods are suitable for different cases. For the short transfer link, the method of optical delay is preferable, because a high sampling rate of the electrical delay method is required to generate the short
time delay. However, for the long transfer link, the optical delay will introduce high attenuation on the transferred optical frequency signals, which makes the electrical delay as a better choice. In the practical application, the two methods can be adopted simultaneously, where the electrical delay generates the coarse time delay and the optical delay tunes the fine time delay.

In conclusion, we proposed a TD-LTW optical frequency comparison technique, which can enhance the phase noise reduction in the localized optical two-way optical frequency comparison. The proposed scheme reserves the advantage of the LTW optical frequency comparison scheme, which can accomplish the data acquisition at one site, guaranteeing the synchronization of data acquisition. This capability also promises the high-performance, real-time frequency comparison technique for the two distant optical signals rather than post processing as demonstrated in the TWC optical frequency comparison scheme \cite{Anthony2014Two,Stefani2015Tackling} and the optical frequency transfer scheme \cite{C2015Doppler}. The experimental results demonstrate that the measured stability of optical frequency comparison is improved from $1.30\times 10^{-15}$ to $5.25\times 10^{-16}$ at the 1 s integration time. In the future work, we are planning to integrate the proposed TD-LTW optical frequency comparison scheme into the bus or ring topology structures to provide the potential of optical clock comparison over large area with a higher stability.

\ifCLASSOPTIONcaptionsoff
  \newpage
\fi



%\bibliographystyle{IEEEtran}

\end{document}